\documentclass[aps,pre,amssymb]{revtex4-2}
\usepackage{graphicx}
\usepackage{color}
\usepackage[utf8]{inputenc}
\usepackage{braket}
\usepackage{graphicx}

\begin{document}

\title{Effect of two vaccine doses in the SEIR epidemic model using a
stochastic cellular automaton}

\author{Enrique C. Gabrick$^1$\footnote{ecgabrick@gmail.com}, Paulo R. Protachevicz$^{2}$, Antonio M.
Batista$^{1,3}$, Kelly C. Iarosz$^{4,5}$, Silvio L. T. de Souza$^6$, Alexandre C.
L. Almeida$^7$, Jos\'e D. Szezech Jr$^{1,3}$, Michele Mugnaine$^8$, Iber\^e L.
Caldas$^2$}
\affiliation{$^1$Postgraduate Program in Sciences, State University of Ponta Grossa,
84030-900, Ponta Grossa, PR, Brazil}
\affiliation{$^2$Physics Institute, University of S\~ao Paulo, 05508-090, S\~ao
Paulo, SP, Brazil}
\affiliation{$^3$Department of Mathematics and Statistics, State University of
Ponta Grossa, 84030-900, Ponta Grossa, PR, Brazil}
\affiliation{$^4$Faculdade de Tel\^emaco Borba, FATEB, Tel\^emaco Borba, PR, Brazil}
\affiliation{$^5$Graduate Program in Chemical Engineering Federal Technological
University of Paran\'a, Ponta Grossa, PR, Brazil}
\affiliation{$^6$Federal University of S\~ao Jo\~ao del-Rei, Campus Centro-Oeste,
35501-296, Divin\'opolis, MG, Brazil}
\affiliation{$^7$Statistics, Physics and Mathematics Department, Federal University
of São Jo\~ao del-Rei, Ouro Branco, MG, Brazil}
\affiliation{$^8$Department of Physics, Federal University of Paran\'a, Curitiba,
PR, Brazil}

\begin{abstract}
In this work, to support decision
making of immunisation strategies, we propose the inclusion of two vaccination
doses in the SEIR model considering a stochastic cellular automaton. We analyse
three different scenarios of vaccination: $(i)$ unlimited doses, $(ii)$ limited
doses into susceptible individuals, and $(iii)$ limited doses randomly
distributed overall individuals. Our results suggest that the number of
vaccinations and time to start the vaccination is more relevant than the vaccine
efficacy, delay between the first and second doses, and delay between vaccinated
groups. The scenario $(i)$ shows that the solution can converge early to a
disease-free equilibrium for a fraction of individuals vaccinated with the first
dose. In the scenario $(ii)$, few two vaccination doses divided into a small
number of applications reduce the number of infected people more than into many
applications. In addition, there is a low waste of doses for the first
application and an increase of the waste in the second dose. The scenario
$(iii)$ presents an increase in the waste of doses from the first to second
applications more than the scenario $(ii)$. In the scenario $(iii)$, the total
of wasted doses increases linearly with the number of applications. Furthermore,
the number of effective doses in the application of consecutive groups decays
exponentially overtime.

\end{abstract}

\maketitle
\section{Introduction}

Understanding the dynamics of epidemics is an important interdisciplinary
research topic \cite{silvio}. The studies can provide contribution to the
prevention and control of infectious diseases \cite{automata}. An epidemic is
the fast spread of infectious diseases that produce many infected individuals
within a population \cite{vac-ca}. Some examples of epidemics are the bubonic
plague or Black Death during the fourteenth century \cite{history}, Spanish flu
in 1918 \cite{spanish}, Severe Acute Respiratory Syndrome (SARS) in 2002
\cite{sars}, H1N1 in 2009 \cite{h1n1}, and, more recently, in the end of 2019,
the novel coronavirus (COVID-19) arose in Hubei Province in China \cite{lancet}.

There are many ways that can be used to control the infectious disease spread,
for instance the minimisation of the social contact \cite{mello}, quarantine
\cite{seir-vac1}, restrictions \cite{silvio}, lockdown \cite{sharma21}, and
others \cite{automata}. One of the most effective strategies is the application
of vaccines \cite{seirv}. Most mathematical models consider only one dose
\cite{seir-vac1, seirv}. For some diseases, there are vaccines that are
administered in two doses, for instance, for the novel coronavirus \cite{seirv}.
With this in mind, we include two new compartments in the SEIR model to
simulate two doses of vaccination. Our SEIR model is described by a stochastic
cellular automaton and can be adapted for many diseases.

Mathematical models have been used to analyse the dynamical behaviour of
epidemics and evaluate strategies to control them
\cite{piccirillo21,amaku21a,amaku21b}. In general, the population is separated
into compartments in the mathematical models of infectious diseases, such as
SIS, SIR, SEIR, and SEIRS \cite{rbef}. The compartments can be susceptible
($S$), exposed ($E$), infected ($I$), and recovered ($R$) individuals. In this
work, we choose the SEIR model \cite{reports}, which can be studied from
differential equations \cite{reports} or cellular automata approach.
Nevertheless, our model can be modified for the SIR model.

In the SEIR model, the host population is divided into four compartments
\cite{seir}: $S$ (susceptible) represents the individuals who can be
infected, when in contact with infected individuals, $E$ (exposed) are the
individuals in latent \cite{sharma21, rbef} and/or incubation period
\cite{amaku21b, quanxing}. The latent period corresponds to the range time in
which the individuals do not transmit the disease \cite{rbef}. 
On the other hand, during the incubation period, the exposed  individuals can transmit the disease with a lower incidence than infected individuals
\cite{amaku21b}. In our simulations, $E$ corresponds to a latent period. $I$
(infected) is associated with the individuals that are infected and can transmit
the disease. $R$ (recovered) is related to the individuals that get
immunity or die. The scenario studied does not consider the possibility of reinfection. This model was considered to simulate the impact of easing
restrictions \cite{silvio}, scenarios with reinfection \cite{reinfeccao},
inclusion of vaccine \cite{heliyon}, spatiotemporal evolution of epidemics
\cite{automata}. More recently, Sharma et al. \cite{sharma21} considered a
SEIRD model (Death) with delay to predict the evolution of pandemic in India.
They analysed scenarios with no lockdown, strict lockdown, and movement with
social distancing.

Etxeberria-Etxaniz et al. \cite{seir-vac} considered the vaccination of newborns
and periodic impulse vaccination in the SEIR model. A stochastic formulation of
the SEIR model with the inclusion of vaccination was studied by Balsa et al.
\cite{seir-vac1}. They considered the combination of vaccination and quarantine.
Jadidi et al. \cite{wireless} proposed the vaccination in two steps, vaccine
allocation, and targeted vaccination. The SEIR model with quarantine, isolation,
and imperfect vaccine was analysed by Safi and Gumel \cite{seir-vac2}. Yongzhen
et al. \cite{seir-vac3} reported the effect of constant and  pulse vaccination
on the SIR model with an infectious period. In the SIR model, White et al.
\cite{vac-ca} included vaccine using cellular automata. Nava et al. \cite{nava}
studied how controllable parameters can lower the infection spread in an open
crowed space. They considered the generalised SEIR model with an analytical and
cellular automaton approach.

Most of the works considered one dose of vaccine. Recently, De la Sen et al.
\cite{two-doses} proposed a SEIR discrete model with two vaccination doses that
are applied in susceptible individuals. They discussed the influence of the
vaccination starting time, as well as the effect of the delay between the first
and second doses. 

In this work, we propose an epidemic SEIR model with two doses vaccinations
based on stochastic cellular automata (CA) \cite{michele, physica,swarm}.
In the deterministic context, Wolfram defined the CA as discrete
idealisations \cite{wolfram2,wolfram1}. Deterministic CA evolves in accordance
with deterministic transition rules \cite{physica,wolfram}, while the
stochastic evolves in accordance with stochastic transition rules \cite{stc1}.
Furthermore, the transition rules can be a mix of deterministic and
probabilistic ones \cite{borges2015,pramana}. One advantage of the CA is the
possibility of including local features in the model \cite{bin}. CA have been
applied in various areas, for instance physics \cite{vichniac84} and biology
\cite{luca21,viana14}. Santos et al. \cite{santos09} reported that a CA model
can reproduce time series of dengue epidemics. Recently, Blavatska and Holovatch
\cite{blavatska21} investigated infection spreading processes in a CA where only
a fraction of individuals is affected by a disease. Mikler et al. \cite{mikler}
studied a stochastic CA to simulate a SIR model with geographic and demographic
characteristics, as well as migratory constraints. In addition,
Cavalcante et al. studied a SEIR model by means of differential equations and
CA \cite{cavalcante}.

We consider three scenarios of vaccination: $(i)$ unlimited doses in susceptible
individuals by means of continuous vaccination, $(ii)$ limited doses in
susceptible individuals through pulse periodic vaccination, and $(iii)$ limited
doses that are randomly distributed. In the scenario $(i)$, we obtain the
variation of the infected individual numbers as a function of the time of
starting the application, the delay between the first and second doses, and the
vaccine efficacy influence. The scenario $(ii)$ permits to find effective ways
to manage few doses. In the scenarios $(ii)$ and $(iii)$, individuals are
vaccinated and the effect occurs only in the susceptible individuals. For these
scenarios, we estimate the occurrence of wasted doses, applied to individuals
who received the first dose and are infected before the second dose, as a
function of the considered control parameters. We show that the waste in the
first dose occurs when the available doses are bigger than the number of
susceptible individuals or the first dose is applied in the individuals outside
the susceptible state (third scenario). Furthermore, we show that the number of
vaccinations and time to start the vaccination is more relevant than the vaccine
efficacy, delay between first and second dose, and delay between vaccinated
groups. In the scenario $(ii)$, few two vaccination doses divided into a small
number of applications reduce the quantity of infected people more than into
many applications. In addition, there is a low waste of doses for the first
application and an increase of the waste in the second dose. The scenario
$(iii)$ presents an increase in the waste of doses from the first to second
applications more than the scenario $(ii)$. In the scenario $(iii)$, the total
of wasted doses increases linearly with the number of applications. The number
of effective doses in the application of consecutive groups decays exponentially
overtime.

The paper is organised as follows. In Section $2$ the model is presented in
details. In Section $3$, numerical simulations of the scenario $(i)$ are
discussed. Section $4$ shows numerical simulations of the scenerii $(ii)$ and
$(iii)$. Finally, Section $5$ presents our conclusions.


\section{Description of the model}

Cellular automata are mathematical models characterised by discrete time, space,
and state variables. In our model, each time step is taken to be one day. The
transitions between states occur by local rules. In this work, we build a
two-dimensional CA with deterministic and probabilistic transition rules. The
CA is given by a lattice ($L$) composed of $N\times N$ identical cells
\cite{ilachinski} with
\begin{equation}
L={(i, j), \ \ i,j\in Z^*_+ \ | \ \ 1 \leq i,j \leq N}.
\end{equation}
Each cell is identified by one state $x(i,j,t)\in U$, where the set
$U=\{1,2,3,4\}$ indicates the $\{S$, $E$, $I$, $R\}$ states, respectively. For
each cell $(i,j)$, a Moore's neighbourhood $V$ is considered and given by
$M(i,j)=(i,j)+v$, where $v\in V$ and $V$ is defined as
\begin{equation}
V = \{(0,0), (-1, 0), (-1,-1),(-1,1), (1,0), (1,-1), (1,1), (0,-1), (0,1)\}.
\end{equation}
The boundary conditions represented by $(i,j)\not\in L$ \cite{vac-ca} are given
by $x(i,j,t)=0$. With this boundary condition the cells are disposed in a plan.
The cells inside of the region delimited by $N \times N$ do not interact with
the boundary cells. One time step (one day) is defined when all cells in the
lattice are updated in accordance with transition rules.

We consider parameters based on various sources in the literature. The values of
the transmission rate are obtained from Balsa et al. \cite{seir-vac1}. From
Radulescu \cite{reports}, we use the values related to the infectious period,
time to develop symptoms, and infection rate. The time between doses is obtained
from De la Sen et al. \cite{two-doses} and the vaccine efficacy from Voysey et
al. \cite{lancet}.

\subsection{Transmission model without vaccination}

\begin{figure}[hbt]
\begin{center}
\includegraphics[scale=0.50]{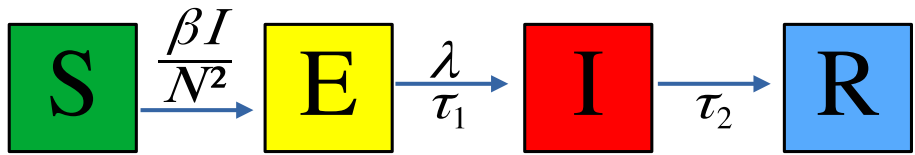}
\caption{Schematic representation of the SEIR model. $\beta$ is the probability
of the susceptible individuals become exposed and $\lambda$ is the rate of
exposed individuals become infected. $\tau_1$ and $\tau_2$ correspond to the
time intervals in which exposed and infected individuals remain in the same
state, respectively.}
\label{fig1}
\end{center}
\end{figure}

The total population $N\times N$ is separated into four compartments, as shown
in Fig. \ref{fig1}. The susceptible, exposed, infected, and recovered cells are
represented by $S(t)$, $E(t)$, $I(t)$, and $R(t)$, respectively. Individuals in
the compartmental $S$ can become $E$ with probability $\beta$ when there are
$I$ neighbours. After $\tau_1$ times, there is a rate
$\lambda$ in which individuals go from $E$ to $I$.  They stay during $\tau_2$
time step in $I$ (infectious period) and then go to $R$. In the model, we
consider that recovered cells have a permanent immunity. The initial condition
is given by a random distribution of infected cells $I(0)$. The major part of
the cells is initially in the susceptible state. The exposed and recovered
cells are not considered in the initial time of the simulation.

The update rule of the individual $(i,j)$ in absence of vaccination depends on
its initial state and their neighbouring states in the previous time, as well
as on the transition probability $\beta$ and $\lambda$ from the exposed to
infected states, and the time constants $\tau_1$ and $\tau_2$. The evolution of
the CA can be expressed by a time function $F$ given by 
\begin{equation}
F(i,j,t)=f(x(i,j,t-1),...,x(i,j,t-\tau), \ x(i+\alpha,j+\gamma,t-1),\beta,
\lambda),
\label{F}
\end{equation}
where $(\alpha,\gamma)\in V$, $\tau\in(\tau_1,\tau_2)$.

The transition rules can be summarised as:
\begin{itemize}
\item Each infected cell ($x(i,j,t)=3$) can infect a neighbour in a
susceptible state ($x(i,j,t)=1$) with probability $\beta$. In other words, if
the cell is in the susceptible state with one or more (up to eight) infected
neighbours, each infected cell will try to transmit the disease to the
susceptible with a probability $\beta$. Once infected, the susceptible cell
($x(i,j,t)=1$) will evolve in the next step (day) to the exposed state
($x(i,j,t)=2$).
\item After evolving to the exposed state ($x(i,j,t) = 2$), the cells stay in
this state by $\tau_1$ time steps (days units). In the next step, a fraction
$\lambda$ of exposed cells evolve to the infected state ($x(i,j,t)=3$).
\item If the cells are infected ($x(i,j,t) = 3$), they remain in the infected
state by $\tau_2$ time steps. After $\tau_2$, infected cells evolve to a
recovered state ($x(i,j,t)=4$). 
\item After evolving to the recovered state ($x(i,j,t)=4$), the cells stay in
there all the time.
\end{itemize}

\begin{figure}[hbt]
\begin{center}
\includegraphics[scale=0.45]{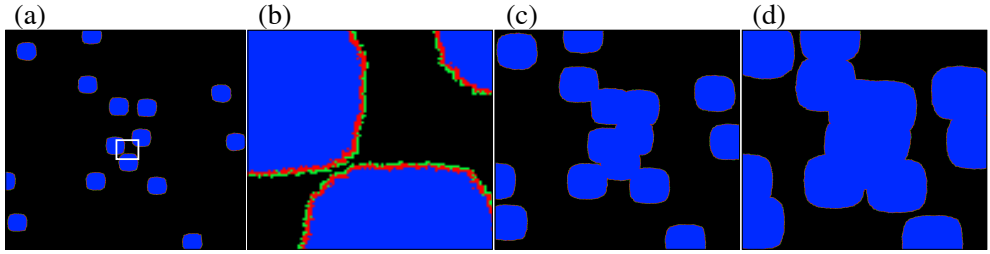}
\caption{Spatial time evolution of the CA for $N=1000$, $I(0)=15$,
$\lambda=1/3$, $\beta=1/4$, $\tau_1=6$, $\tau_2=14$. (a) States distribution for
$t=300$ and (b) magnification of the region delimited by the white square in
panel (a). The panels (c) and (d) show the states distribution for $t=600$ and
$t=900$, respectively. In the black, red, green, and blue regions, we observe
the susceptible, infected, exposed, recovered states, respectively.}
\label{fig2}
\end{center}
\end{figure}

An illustration of one spatial time evolution of the CA is displayed in Fig.
\ref{fig2}.  In Fig. \ref{fig2}(a), we see the states distribution for $t=300$.
Figure \ref{fig2}(b) exhibits the magnification of the region delimited in the
panel (a) by a white square. Figures \ref{fig2}(c) and \ref{fig2}(d) show the
spatial evolution for $t=600$ and $t=900$, respectively. In Fig. \ref{fig2}(b),
the borders composed of green and red cells represent the individuals in the
exposed and infected states, respectively. These borders evolve overtime in
circular waves from the central region, as shown by means of the blue cells that
represent the recovered states. In the final evolution of the system, the
lattice exhibits only blue cells, which is the disease-free equilibrium
solution, where the eradicating of the illness is found.

\subsection{Transmission model with vaccination}

\begin{figure}[hbt]
\begin{center}
\includegraphics[scale=0.4]{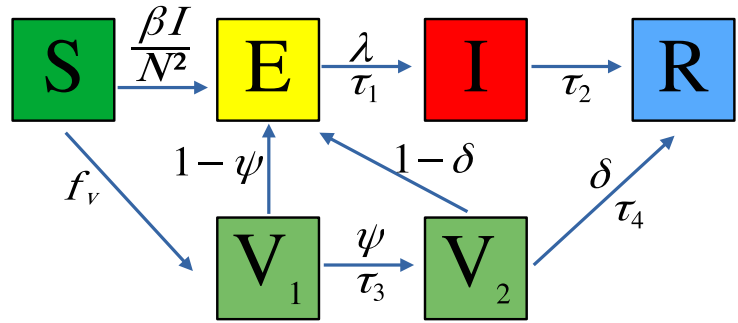}
\caption{SEIR model with two vaccination doses. The first and second doses are
the two new cell states, being represented by $V_1$ and $V_2$. The parameters
$f_{\rm v}$ define the fraction of the susceptible that receive the first dose.
$\psi$ and  $\delta$ correspond to the first and second doses efficacy,
respectively. $\tau_3$ is the time interval between the first and second doses,
and $\tau_4$ is the time to arrive in the recovered state from the $V_2$.}
\label{fig3}
\end{center}
\end{figure}

In our CA model, we include vaccinations that are divided into unlimited and
limited doses. In this case, we consider the states $U=\{1,2,3,4,5,6\}$, where
the two new ones $x(i,j,t)=5$ and $x(i,j,t)=6$ represent the cells that receive
the first $V_1$ and second $V_2$ doses of the vaccine. Figure \ref{fig3}
displays the schematic representation of the SEIR model with the first and
second doses, named as SEIR2V. This schematic representation is valid for the
scenario $(i)$ (unlimited doses) and scenario $(ii)$ (limited doses), where only
the susceptible cells are vaccinated. In the scenario $(iii)$, the vaccine doses
are randomly distributed in all lattice. 

\begin{figure}[hbt]
\begin{center}
\includegraphics[scale=0.5]{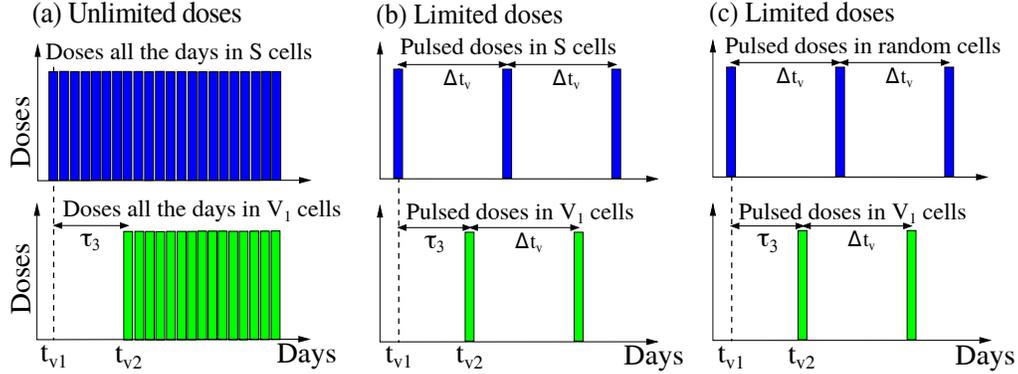}
\caption{Schematic representation of three scenarios for vaccination. The panel
(a) illustrates the $(i)$ scenario. The panel (b) is a schematic representation
of the $(ii)$ scenario, and the panel (c) is the representation of the $(iii)$
scenario. The blue bars indicate the amount for first dose and the green bar
correspond to the second dose.}
\label{vacinas}
\end{center}
\end{figure}

The scenarios for vaccine application are illustrated in Fig. \ref{vacinas}. The
blue bars indicate the total available first doses and the green bars
correspond to the second doses. First, we consider the scenario $(i)$, as
illustrated in Fig. \ref{vacinas}(a). In the time $t_{\rm v_1}$, a fraction
$f_{\rm v}$ of susceptible cells are vaccinated with the first dose, 
evolving to the $V_1$ state. In the time $t_{\rm v_1}+1$, another susceptible
fraction ($f_{\rm v}$) receives the first dose. And so on, every day, until the
infected number goes to zero. In this sense, we say that available doses are
unlimited. The cells that go to the first dose vaccinated state $V_1$, remain
in this state during $\tau_3$ time steps before receiving the second dose.
The time $\tau_3$ is the delay between the first and second doses.
The probability of these cells get infected is $1-\psi$, where $\psi$ is the
first dose efficacy. After $\tau_3$ (time delay), in $t_{\rm v2}$, the second
dose is applied in the cells in $V_1$ states. The cells, that get the second
dose, are the cells that were not infected while stay in $V_1$ state. After
receiving the second dose, the probability of the cells in $V_2$ states get
infected is $1-\delta$, where $\delta$ is the booster efficacy. The cells, that
are not infected in the period $\tau_4$, evolve to $R$ state. Fig.
\ref{vacinas}(b) illustrates the $(ii)$ scenario, where the amount
of dose is a fixed number, given by $D_{\rm T}=f_{\rm v}\cdot N^2$. The
vaccination starts at $t_{\rm v_1}$ with the application of $D_{\rm T}$ doses in
the susceptible cells. The next group receives the first dose after
$\Delta t_{\rm v}$ times in a pulsed way. The application of the second dose
occurs in the same way. The second dose application starts in $t_{\rm v2}$ with
$D_{\rm T}$ doses that will be applied in uninfected cells belonging to the first
vaccinated group. The different susceptible groups and $V_1$ groups, that
receive the first and second doses, are denoted by $m$ with $m\in Z^*_+$. The
time application of the first and second doses in subsequently cell groups is
given by $t^m_{\rm v_{1,2}}=t^{m-1}_{\rm v_{1,2}}+\Delta t_v$, where $m\geq 2$,
$t^1_{\rm v_{1,2}}=t_{\rm v_{1,2}}$, and $\Delta t_{\rm v}$ is the interval between the
applications. In this context, the $(i)$ scenario has $\Delta t_{\rm v} = 1$.
Fig. \ref{vacinas}(c) illustrates the $(iii)$ scenario. The application
protocols are the same that $(ii)$ scenario. The difference is that instead of
applying the first dose in susceptible cells, the application of $D_{\rm T}$
doses occurs in all cells that belong to the lattice. However, the effect occurs
only in susceptible cells. In this sense, we use the term wasted doses to refer to the doses given to exposed, infected, or recovered individuals. These applied doses will not contribute to preventing newly infected individuals.
In this way, the total number of doses is represented by
$N_{\rm T}=D_{\rm T}\cdot N_{\rm App}$, where $N_{\rm App}$ is the total number of
applications. The total amount of vaccines in each dose $D_{\rm T}$ can be
identified as the sum of the  effective ($D_{\rm eff}$) and wasted doses
($D_{\rm w}$), corresponding to $D_{\rm T}=D_{\rm eff}+D_{\rm w}$.

To simulate the CA with the two new compartments, we consider new rules:
\begin{itemize}
\item Given a cell in $V_1$ state, if this cell has one or more infected
neighbours, it can be infected with $1-\psi$ probability, where $\psi$ is the
first dose efficacy. If the vaccinated cell is infected, it evolves to an
exposed state, such as a susceptible cell in the model without vaccination.
\item The uninfected cells in $V_1$, evolve to $V_2$ state after a time
delay, $\tau_3$.
\item If the cell in $V_2$ has one or more neighbours in an infected state,
it can be infected with $1-\delta$ probability, where $\delta$ is the booster
efficacy. If the vaccinated cell is infected, it evolves to an exposed state.
The cells that cannot be infected evolve to the recovered states after $\tau_4$.
\end{itemize}

\begin{figure}[hbt]
\begin{center}
\includegraphics[scale=0.6]{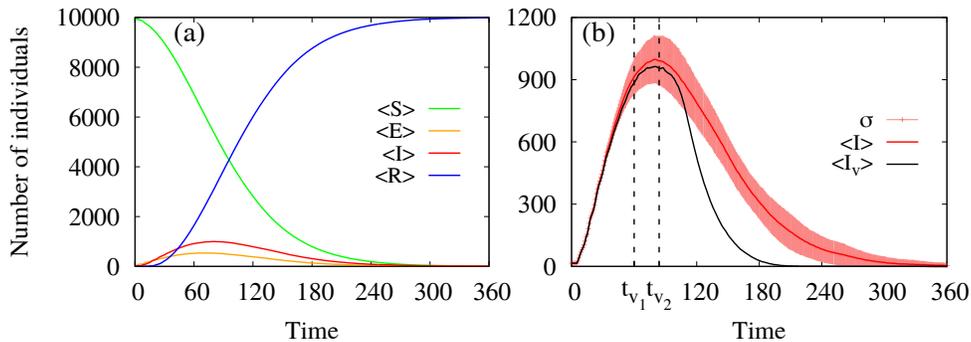}
\caption{(a) Average time of the SEIR model without vaccination. The green,
orange, red, and blue lines denote $\langle S\rangle$, $\langle E\rangle$,
$\langle I\rangle$, and $\langle R\rangle$, respectively. (b) Number of infected
individuals for the case without vaccination (red line), where $\sigma$ is the
standard deviation for $50$ simulations, and for the scenario $(i)$ with
vaccination (black line). We consider $\beta=1/4$, $\lambda=1/3$, $\tau_1=6$,
$\tau_2=14$. In the panel (b), the parameters of the vaccination are
$t_{\rm v_1}=60$, $t_{\rm v_2}=84$ (indicated by vertical dashed lines),
$\tau_3=24$, $\tau_4=10$, $f_{\rm v}=0.02$, $\psi=0.66$, and $\delta=0.75$. The
time unit is day.}
\label{fig4}
\end{center}
\end{figure}

Figure \ref{fig4}(a) shows the time evolution of the SEIR model without vaccine
application, where the green, orange, red, and blue lines represent the average
of susceptible, exposed, infected, and recovered cells, respectively. In Fig.
\ref{fig4}(b), we present the infected curve for the scenario $(i)$ of
vaccination (black line) and no vaccine (red line). In the considered example,
the first and second vaccination times are, respectively, $t_{\rm v_1}=60$ and
$t_{\rm v_2}=84$. With and without vaccination control, the disease eradication is
obtained for $T_0=218$ and $360$ time steps, respectively.

In Sections 3 and 4, we consider $N=100$, $\lambda=1/3$, $\beta=1/4$,
$\tau_1=6$ days, $\tau_2=14$ days, $\tau_3=24$ days, $\tau_4=10$ days. The
initial numbers of cells are given by $I(0)=15$, $S(0)=N^2-I(0)$, and
$E(0)=R(0)=V_1(0)=V_2(0)=0$. Our results are calculated by means of an average
of $50$ independent repetitions, denoted by $\langle .\rangle$.


\section{Unlimited doses}

\begin{figure}[hbt]
\begin{center}
\includegraphics[scale=0.6]{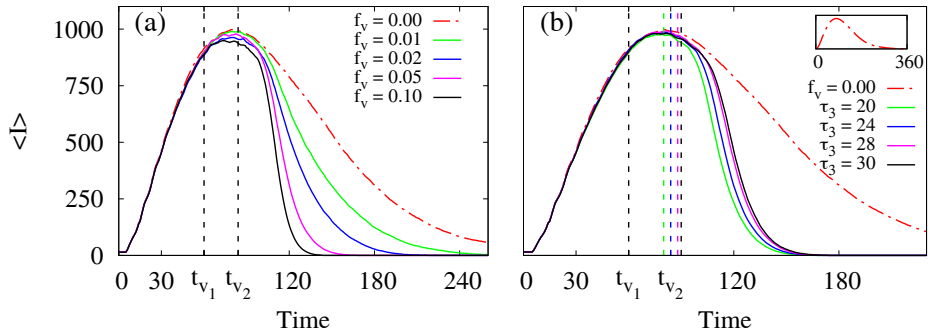}
\caption{(a) Average of infected cells overtime for different values of
$f_{\rm v}$. Vertical black dashed lines indicate the starting of the first
($t_{\rm v_1}=60$) and second ($t_{\rm v_2}=84$) doses. In the panel (b), infected
cells overtime for different values of $\tau_3$ and $f_{\rm v}=0.05$. The
vertical coloured lines represent the starting of the second dose vaccination
for different values of $\tau_3$. The red dash-dotted line represents the case
without vaccination. In the panel (b), the insert shows the whole curve. The
considered value of vaccination efficacy for the first and second doses are
$\psi=0.66$ and $\delta=0.75$. The time unit is day.}
\label{fig5}
\end{center}
\end{figure}

Figures \ref{fig5}(a) and \ref{fig5}(b) exhibit the infected cells overtime due
to the influence of different values of $f_{\rm v}$ and $\tau_3$, respectively.
The red dash-dotted line represents the case without vaccination. The vertical
black dashed line indicates the start of the first and second doses. In Fig.
\ref{fig5}(a), the infected curve suffers narrowing when $f_{\rm v}$ is
increased, implying in the decrease of the area under the respective curve. The
normalised area ($A_{\rm n}=1$) covers the red dash-dotted curve. Increasing the
$f_{\rm v}$ values from zero, the normalised areas are equal to $A_{\rm n}=0.80$,
$0.70$, $0.62$, and $0.57$, respectively. As a result, the system goes to the
disease-free equilibrium point earlier. In Fig. \ref{fig5}(b), we observe that
an increase in $\tau_3$ generates wider infected curves. The green, blue,
magenta, and vertical black dashed lines indicate the start time of the second
dose considering $\tau_3=20$, $24$, $28$, and $30$, corresponding to
$A_{\rm n}=0.59$, $0.62$, $0.65$, and $0.65$, respectively. For $\tau_3\geq 20$
occurs an increase in the area, however, the influence of this parameter on the
dynamics is less pronounced than $f_{\rm v}$.

\begin{figure}[hbt!]
\begin{center}
\includegraphics[scale=0.6]{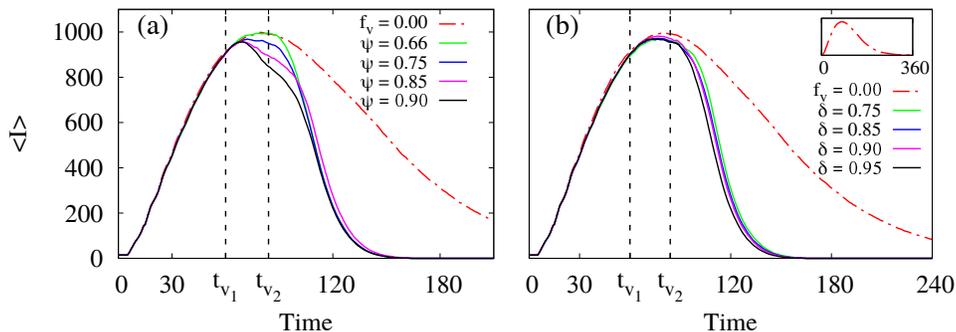}
\caption{(a) Average of infected cells overtime for different values of $\psi$
for $\delta=0.95$ and $f_{\rm v}=0.05$. (b) Different values of $\delta$ overtime
for $\psi=0.66$ and $f_{\rm v}=0.05$. The vertical dashed lines indicate the time
in which the first and second doses are administrated. The red dash-dotted line
corresponds to the case without vaccination. The time unit is day.}
\label{fig6}
\end{center}
\end{figure}

We investigate the influence of the efficacy of the first and second doses,
denoted by $\psi$ and $\delta$, respectively, in the infected curves. Figures
\ref{fig6}(a) and \ref{fig6}(b) display the infected curves overtime for
different values of $\psi$ and $\delta$. In Fig. \ref{fig6}(a), we see that the
infected curve goes faster to zero for large values of $\psi$. In this case, we
consider the second dose efficacy as $\delta=0.95$. For $\psi=0.66$, Figure
\ref{fig6}(b) shows that the infected curve is narrowed when $\delta$ is
increased. In both cases, the areas do not change significantly. Therefore, for
this range of values, we do not observe significant effect on the dynamical
system. 

\begin{figure}[hbt]
\begin{center}
\includegraphics[scale=0.6]{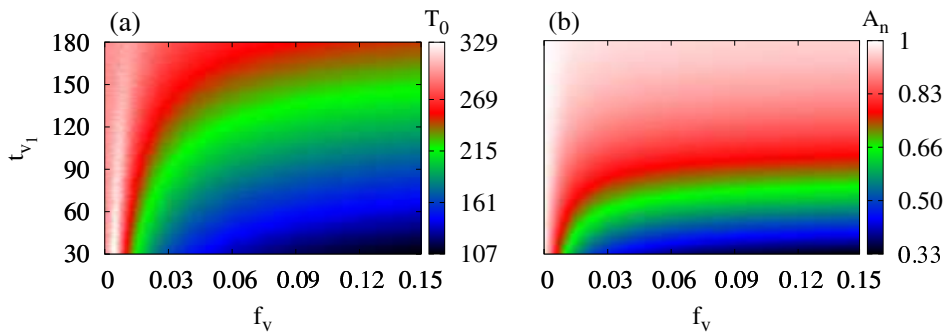}
\caption{(a) Time for the disease eradicating ($T_0$) and (b) normalised area
under the infected curve ($A_{\rm n}$) as a function of $t_{\rm v_1}$ and
$f_{\rm v}$ for $\psi = 0.66$ and $\delta = 0.75$. The time unit is day.}
\label{fig7}
\end{center}
\end{figure}

Figure \ref{fig7}(a) exhibits the time for disease eradication ($T_{\rm 0}$) as a
function of $t_{\rm v_1}$ and $f_{\rm v}$. In Fig. \ref{fig7}(b), we calculate the
normalised area ($A_{\rm n}$) under the infected curve as a function of
$t_{\rm v_1}$ and $f_{\rm v}$. Our results show that earlier to start the first
dose application more effective is the vaccine. The $f_{\rm v}$ values larger
than $0.01$ affect positively the dynamics for $t_{v_1}\le 30$, namely, for
these values $T_0$ and $A_{\rm n}$ are significantly reduced. For small
$f_{\rm v}$ or large $t_{\rm v_1}$ values, the effect of vaccination on $T_0$ can
occur, however, it is reduced. The major reduction in $T_0$ and $A_{\rm n}$
happens for large $f_{\rm v}$ and small $t_{\rm v_1}$. For $f_{\rm v} \leq 0.01$, the
effects of $T_0$ are smaller than $f_{\rm v}\approx 0.01$. The total number of
infected individuals is smaller for $f_{\rm v} \geq 0.01$, as shown in Fig.
\ref{fig7}(b).

\section{Limited doses}

\begin{figure}[hbt]
\begin{center}
\includegraphics[scale=0.6]{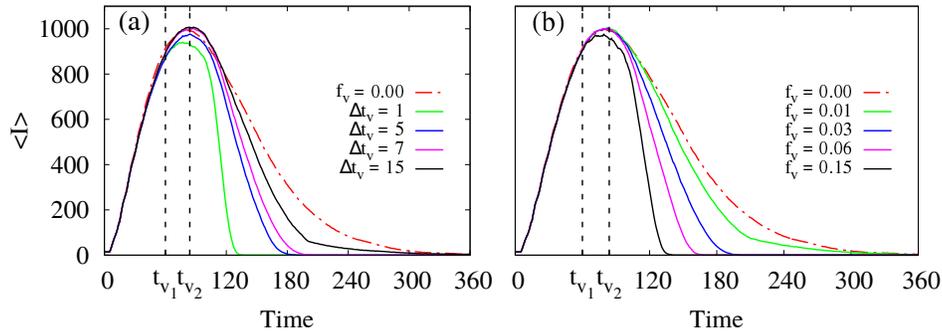}
\caption{(a) Average of infected cells overtime for different values of
$\Delta t_{\rm v}$ and $f_{\rm v}=0.03$. The green, blue, magenta, and black lines
correspond to the values of $\Delta t_{\rm v}$ equal to $1$, $5$, $7$, and $15$,
respectively. (b) Average of infected cells overtime for different values of
$f_{\rm v}$ and $\Delta t_{\rm v}=7$. The green, blue, magenta, and black curves
correspond to the values of $f_{\rm v}$ equal to $0.01$, $0.03$, $0.06$, and
$0.15$, respectively. The vertical dotted lines indicate the first and second
doses applied in the first cell groups. The red dash-dotted line represents the
situation without vaccination. The parameters for vaccination are given by
$\psi=0.66$, $\delta=0.75$, $t_{\rm v_1}=60$, and $t_{\rm  v_2}=84$. The time unit
is day.}
\label{fig8}
\end{center}
\end{figure}

In the scenario $(ii)$, we investigate the effect of different times among
different groups who received the first and second doses ($\Delta t_{\rm v}$) in
the susceptible cells. We consider $f_{\rm v}=0.03$ and $D_{\rm T}=300$ for each
application, as well as we vary the values of $\Delta t_{\rm v}$, as shown in
Fig. \ref{fig8}(a). The values of $A_{\rm n}$ are equal to $0.58$, $0.73$,
$0.79$, and $0.89$ for the green, blue, magenta and black curves, respectively.
Then, an increase in $\Delta t_{\rm v}$ contributes to an increase in $A_{\rm n}$.
Therefore, the vaccine is more effective for small values of $\Delta t_{\rm v}$.

Figure \ref{fig8}(b) shows the influence of $f_{\rm v}$ for $\Delta t_{\rm v}=7$.
In the scenario $(ii)$, the wasted doses in the first dose vaccination occur
when $D_{\rm T}$ is larger than the number of susceptible individuals. For
$f_{\rm v}=0.01$ (green curve), $f_{\rm v}=0.03$ (blue curve), $f_{\rm v}=0.06$ 
(magenta curve), and $f_{\rm v}=0.15$ (black curve) the cells number who receive
the first dose are $D_{\rm eff}\approx 1500$, $3621$, $4835$, and $6073$,
respectively. The $A_{\rm n}$ values are given by $0.92$, $0.78$, $0.70$, and
$0.60$. For $f_{\rm v}=0.01$, the number of infected individuals goes to zero
approximately when there is no vaccination. For large values of $f_{\rm v}$, the
number of infected individuals goes to zero faster. The increase of $f_{\rm v}$
implies in a reduction of the area and the equilibrium point is achieved
earlier.

\begin{figure}[hbt]
\begin{center}
\includegraphics[scale=0.35]{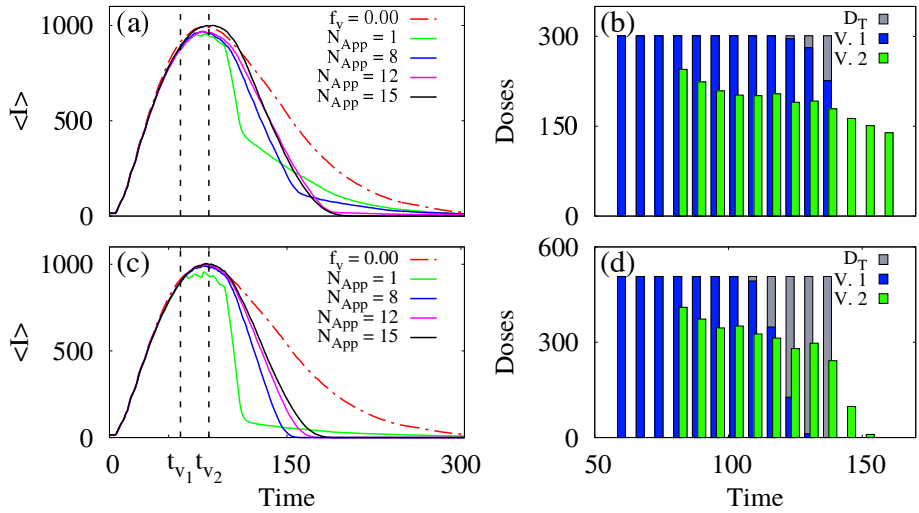}
\caption{(a) Average of infected curve overtime for different values of
$N_{\rm App}$ and $N_{\rm T}=3621$ for each dose. (b) Time distribution of first
and second vaccine doses for $N_{\rm App}=12$ in the panel (a). (c) Average of
infected curve overtime for different values of $N_{\rm App}$ and $N_{\rm T}=6073$
for each dose. (d) Time distribution of vaccine doses for $N_{\rm App}=12$ in the
panel (c). In the infected curves, the green, blue, magenta, and black lines
correspond to $N_{\rm App}=1$, $8$, $12$, and $15$, respectively. The vertical
black dashed lines indicate the time of the first and second doses for the first
cell groups. The red dash-dotted line represents the case without vaccination.
In the time distribution of vaccines, the blue and green bars represent the
first and second effective doses, respectively. We consider $\psi=0.66$,
$\delta=0.75$, $t_{v_1}=60$, $t_{v_2}=84$, and $\Delta t_{\rm v}=7$. The time unit
is days.}
\label{fig9}
\end{center}
\end{figure}

Figure \ref{fig9} displays the effect of dividing the total amount of doses
$N_{\rm T}$ into many applications $N_{\rm App}$ that are spaced out by
$\Delta t_{\rm v}=7$ with $N_{\rm App} \in Z^*$. Figures \ref{fig9}(a) and
\ref{fig9}(b) exhibit the dynamics for $N_{\rm T}=3621$ and the time distribution
of first and second vaccine doses for $N_{\rm App}=12$. The values of $A_{\rm n}$
are equal to $0.74$, $0.77$, and $0.79$. Figure \ref{fig9}(c) shows the time
series for $N_{\rm T}=6073$ and the time distribution of vaccine for
$N_{\rm App}=12$ is shown in \ref{fig9}(d). In Figs. \ref{fig9}(b) and
\ref{fig9}(d), the blue and green bars indicate the effective doses in the first
and second doses applications, respectively. The gray bars denote the total
quantity of available first doses ($D_{\rm T}$). For simplicity, the total
quantity of available second doses is omitted in the figures, however, the same
$D_{\rm T}$ is displaced in time. The effectiveness of the first doses,
represented by the blue bars, is equal to $D_{\rm eff}=3503$ in the panel (b) and
$D_{\rm eff}=4529$ in the panel (d). In the last applications of the first doses,
we observe a decrease in the effective doses. This behaviour occurs due to the
number of available dose ($D_{\rm T}$) to be larger than the number of
susceptible cells. The effectiveness of the second dose, represented by the
green bars, suffers a higher decrease in the amplitude, that occurs by a sum of
factors. The cells that receive the second doses come from the successful first
doses. In this way, it is expected a wasted of at least $1-\psi$ in the $V_2$
state for the range time that $D_{\rm T}$ is not larger than the number of
susceptible individuals. This behaviour can be seen in the time distribution
shown in Fig. \ref{fig9}(b).

In Fig. \ref{fig9}(c), the behaviour significantly depends on $N_{\rm App}$. The
$A_{\rm n}$ values are $0.58$, $0.68$, $0.73$, and $0.75$. Therefore, the same
doses quantity has an effect more pronounced for a small number of applications,
despite the eradication point is reached later. In Fig. \ref{fig9}(d), the blue
bars decrease due to the fact that the number of susceptible individuals is less
than the available doses. In the green bars, the decrease occurs by the same
effect that in Fig. \ref{fig9}(b). The decay is more pronounced by the decrease
in the blue bar. 

\begin{figure}[hbt]
\begin{center}
\includegraphics[scale=0.6]{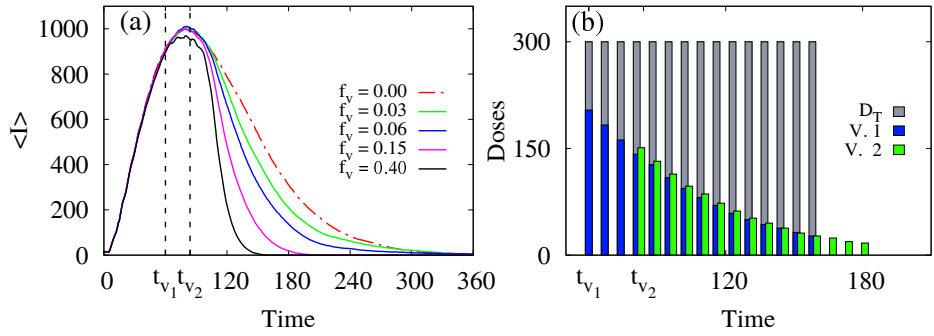}
\caption{(a) Average infected curve overtime for different values of
$f_{\rm v}$. The green, blue, magenta, and black curves denote the results for
$f_{\rm v}$ equal to $0.03$, $0.06$, $0.15$, and $0.40$. The vertical dashed
lines indicate the starting of vaccination. The red dash-dotted line is without
vaccination. (b) Time distribution of the effective doses for $f_{\rm v}=0.03$.
The gray bar indicates the available quantity of each dose ($D_{\rm T}$), the
blue bar indicates the effective first dose ($V_1$), and the green bar denotes
the effective second dose ($V_2$). We consider $\psi=0.66$, $\delta=0.75$, and
$\Delta t_{\rm v}=7$. The time unit is day.}
\label{fig10}
\end{center}
\end{figure}

The scenario $(iii)$ is similar to the scenario $(ii)$, except that the doses
are randomly distributed over all cells. Figure \ref{fig10}(a) shows the
infected cells overtime for some values of $f_{\rm v}$ for $\Delta t_{\rm v}=7$.
Figure \ref{fig10}(b) displays the time distribution of the available vaccine
$D_{\rm T}$ (gray bar), and the effective first (blue bar) and second (green bar)
doses for $f_{\rm v}=0.03$. In Fig. \ref{fig10}(a), the normalised area under the
green, blue, magenta, and black curves are $A_{\rm n}=0.91$, $0.83$, $0.69$, and
$0.59$, respectively. Similar results are found in the scenario $(ii)$
(Fig. \ref{fig8}(b)), namely, the vaccination becomes more effective for large
$f_{\rm v}$. However, the wasted doses in $V_1$ and $V_2$ are larger. This occurs
due to the fact that the cells leave the susceptible state overtime. 

\begin{figure}[hbt]
\begin{center}
\includegraphics[scale=0.6]{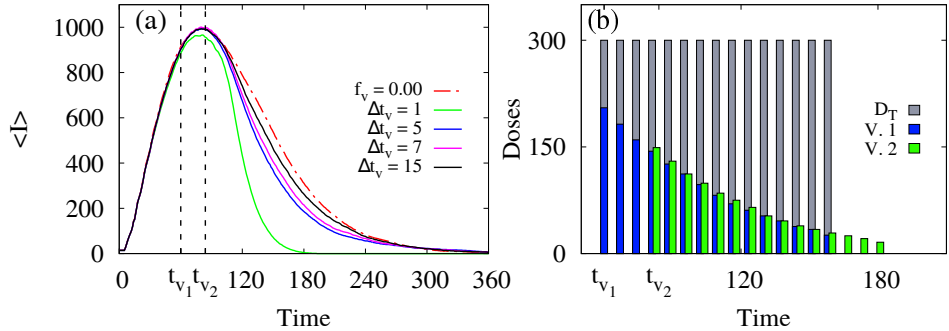}
\caption{(a) Infected cells curves in the scenario $(iii)$ considering different
values of $\Delta t_{\rm v}$. The green, blue, magenta, and black curves
correspond to $\Delta t_{\rm v} = 1$, $5$, $7$, and $15$, respectively. The
vertical black dashed lines indicate the time of application of the first and
second doses at $t_{\rm v_1}=60$ and $t_{\rm v_2}=84$. The red dash-dotted line
corresponds to the case without vaccination. (b) Time distribution for the
vaccine, where the gray bar is the total dose available ($D_{\rm T}$), the blue
bars correspond to the effective dose for the first dose ($V_1$), and the green
bar is the effective dose for the second dose ($V_2$). We consider $\psi=0.66$,
$\delta=0.75$, $f_{\rm v}=0.03$. The time unit is day.}
\label{fig11}
\end{center}
\end{figure}

The influence of $\Delta t_{\rm v}$ on the infected curves is shown in Fig.
\ref{fig11}(a) for $f_{\rm v}=0.03$. The values of $A_{\rm n}$ are equal to
$0.65$, $0.88$, $0.91$, and $0.95$ for the green, blue, magenta, and black
curves, respectively. In Fig. \ref{fig11}(b), the time vaccine distribution 
shows the effective doses for the first $V_1$ (blue bars) and second $V_2$
(green bars) applications, as well as the available doses $D_{\rm T}$ (gray
bars). The fraction of wasted doses is $0.68$ for $V_1$ and $0.78$ for $V_2$. We
verify that the vaccine intervention is more effective for smaller
$\Delta t_{\rm v}$, however, the wasted doses are larger. 

\begin{figure}[hbt]
\begin{center}
\includegraphics[scale=0.3]{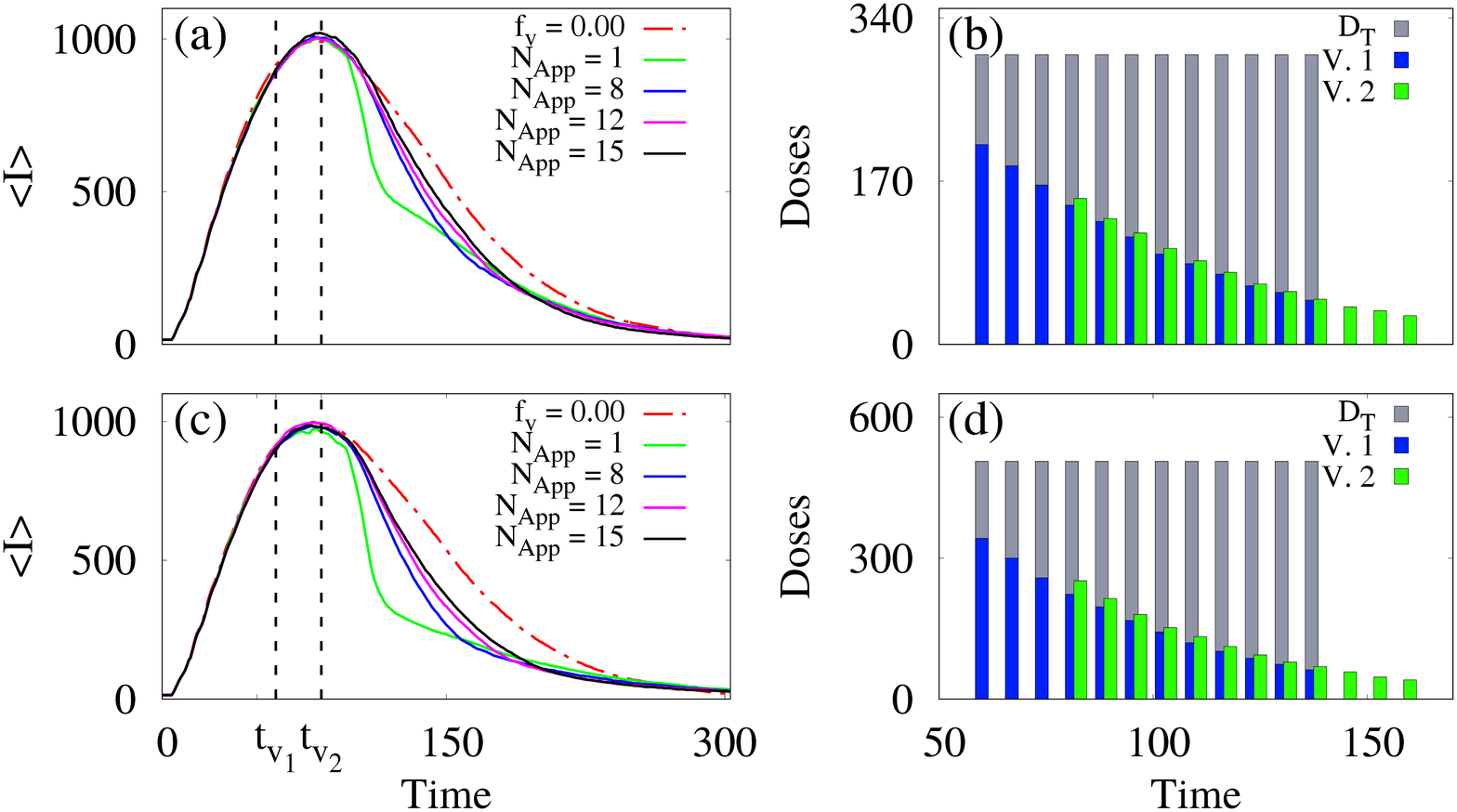}
\caption{(a) Average of infected curves overtime for $N_{\rm T}=3621$ doses and
$N_{\rm App}$ equal to $1$ (green), $8$ (blue), $1$2 (magenta), and $15$ (black).
(b) Time distribution of vaccines for $N_{\rm App}=12$ with the effective vaccine
for the first doses (blue bars), second doses (green bars), and available doses
(gray bar). (c) Average of infected curves overtime for $N_{\rm T}=6073$ doses.
(d) Time distribution of vaccines for $N_{\rm App}=12$ with the effective vaccine
for the first doses (blue bars), second doses (green bars), and available doses
(gray bar). The red dashed-dotted line is without vaccination. The black
vertical dashed lines indicate the time of the first and second doses for the
first cell groups. We consider $\psi=0.66$, $\delta=0.75$, and
$\Delta t_{\rm v}=7$. The time unit is day.}
\label{fig12}
\end{center}
\end{figure}

We investigate the influence of the number of applications considering the
scenario $(iii)$. The results in Fig. \ref{fig12} are similar to Fig.
\ref{fig9} (scenario $(ii)$). Figure \ref{fig12}(a) shows the time evolution of
the infected curve for $3621$ doses and $N_{\rm App}$ equal to $1$ (green), $8$
(blue), $1$2 (magenta), and $15$ (black). The time distribution of the vaccines
for $N_{\rm App} = 12$ is shown in Fig. \ref{fig12}(b). Figures \ref{fig12}(c)
and \ref{fig12}(d) exhibit our results for $6073$ available doses. The effect of
vaccine application is less pronounced than in Fig. \ref{fig9} due to the
number of wasted doses to be larger, as shown in Fig. \ref{fig12}(b). The areas
for these curves increase with the application number. Nevertheless, an increase
in the application number implies an increase in the wasted doses. The effective
doses decrease overtime, as a result of the increase in the number of exposed,
infected and recovered cells overtime. Furthermore, we observe that in the
scenario $(iii)$, the number of effective doses in the application of
consecutive groups decays exponentially according to $\Delta t_{\rm v}$.

\begin{figure}[hbt!]
\begin{center}
\includegraphics[scale=0.7]{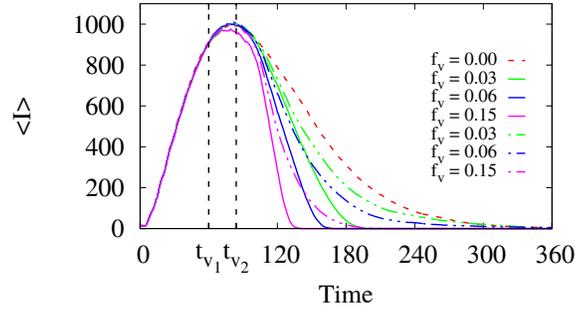}
\caption{Comparison between the scenarios $(ii)$ (continued lines) and $(iii)$
(dash-dotted lines). The vaccinations start at $t_{\rm v_1}=60$ and $t_{\rm v_2}=84$
(vertical dashed lines). The red dashed line is for the case without
vaccination. We consider $\psi=0.66$, $\delta=0.75$, and $\Delta t_{\rm v}=7$.
The time unit is day.}
\label{fig13}
\end{center}
\end{figure}

The difference between the scenarios with limited doses is that for the scenario
$(ii)$ the individuals are tested and then receive the vaccines, while in the
scenario $(iii)$, the vaccines are randomly distributed and consequently it is
observed a larger waste of doses. In Fig. \ref{fig13}, the continuous lines
show the results for the scenario $(ii)$ and dash-dotted lines for the scenario
$(iii)$. These results demonstrate that the scenario $(ii)$ is more effective
than the scenario $(iii)$ when the same number of doses are available.

\begin{figure}[hbt]
\begin{center}
\includegraphics[scale=0.6]{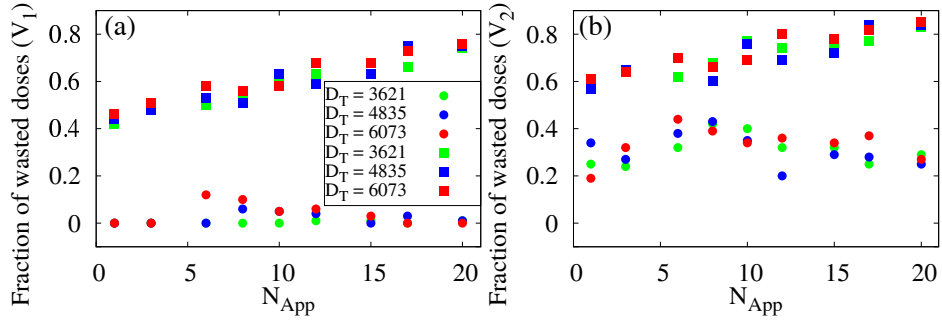}
\caption{Comparison between the fraction of wasted doses as a function of
$N_{\rm App}$ in the scenarios $(ii)$ (circles) and $(iii)$ (squares) for the (a)
first and (b) second doses. We consider $\psi=0.66$, $\delta=0.75$,
$\Delta t_{\rm v}=7$, $t_{\rm v_1}=60$, and $t_{\rm v_2}=84$. The time unit is day.}
\label{fig14}
\end{center}
\end{figure}

Figure \ref{fig14} displays the fraction of wasted dose as a function of
$N_{\rm App}$ for the scenarios $(ii)$ and $(iii)$ by means of circles and
squares, respectively. In some cases occur the waste in the first dose for the
scenario $(ii)$ due to the fact that the total quantity of available doses is
larger than the number of susceptible individuals. In these cases, the average
of wasted doses is equal to $0.01$. In the second dose, the wasted doses in the
scenario $(ii)$ remains approximately constant, exhibiting an average value
about $0.32$. On the other hand, the wasted doses in the scenario $(iii)$ have a
linear dependency on the number of applications. For the first dose, the average
of the wasted doses is about $0.60$ and for second is about $0.72$.

\newpage
\section{Conclusions}

In this paper, we introduce two vaccination doses in the SEIR model by means of
a stochastic cellular automaton, named SEIR2V. We consider probabilistic
transitions in our cellular automaton. There are transitions from susceptible to
exposed states and from exposed to infected states after latent periods. The
individuals in the infected state go to recovered states after an infectious
period. In our model, the disease propagation occurs by contact in a
two-dimensional space with a fix position, such as in a conduction process. A
discussion of convection-like process, i.e. movements in the model, can be
found in \cite{belik} and the influence in a CA model in \cite{weng}. In future
works, we plan to analyse the effects of movement in our model. We use
parameters that are related to types of illnesses and can be adapted for many
other infectious diseases. Depending on the parameter values, it is possible to
find an equilibrium point that is related to the disease-free.

In our model, we consider the inclusion of two new compartments, which are
associated with the individuals vaccinated with the first and second doses. As a
result, a fraction of individuals transit between these two states. The dynamic
behaviour of the model depends on the time between the dose applications,
the efficacy of the first and second doses, and the time of immunisation after
the vaccine application. The inclusion of two doses vaccinations allows us to
investigate the influence of different strategies to realised the immunisation
of the individuals. In this work, we analyse two major scenarios, where there
are unlimited and limited doses of vaccines.

Unlimited doses of vaccines are considered in the scenario $(i)$, while
limited doses are analysed in the scenarios $(ii)$ and $(iii)$. In the scenario
$(i)$, we observe that the cellular automaton converges early to a disease-free
equilibrium for a fraction of individuals vaccinated with the first dose
($f_{\rm v}$) greater than $0.01$ starting at a small time ($t_{\rm v_1}$). Similar
results are found for other $t_{\rm v_1}$ values when $f_{\rm v} \geq 0.08$. For
small values of $t_{\rm v_1}$, earlier eradicating points are achieved. More
important than vaccine efficacy and delay between the first and second doses are
the quantity of effective vaccination and how earlier the application starts.

The scenarios $(ii)$ and $(iii)$ are more realistic, since in real situations
the number of doses is limited. The scenario $(ii)$ represents the case in which
all population is tested and only the susceptible individuals are vaccinated. In
this scenario, we consider a new parameter $\Delta t_{\rm v}$, which is the
interval between the applications of vaccines in different groups. We verify
that the vaccination is more effective for small intervals of applications. When
the total available doses are divided into many applications, the best strategy
is to administrate the doses through few applications. Despite the disease-free
points, they are reached earlier when the number of applications ($N_{\rm App}$)
is larger, while the total number of infected cells is small when $N_{\rm App}$ is
small. In this strategy, the number of wasted doses in the first application is
minimal ($\le13\%$), while the fraction of wasted doses in the second
application is in the range from $20\%$ to $45\%$. 

The scenario $(iii)$ corresponds to the case in which the population is not
tested and the available doses are randomly distributed. In this scenario, the
vaccination is less effective than the scenario $(ii)$ due to the number of
wasted doses. The dose effectiveness decays exponentially according to
$\Delta t_{\rm v}$. The wasted doses can be minimised in this scenario by
collecting the available doses and apply them just once. In this strategy,
the number of wasted doses in the first and second applications exhibits
approximately a linear growth with the number of applications, being larger than
the scenario $(ii)$.

All in all, independently from the strategy, the results can be improved when
the vaccination campaign starts early and with a large number of vaccinated
individuals.


\section*{Acknowledgements}
This work was possible by partial financial support from the following Brazilian
government agencies: Funda\c c\~ao Arauc\'aria, CNPq (407543/2018-0,
302903/2018-6, 420699/2018-0, 407299/2018-1, 428388/2018-3, 311168/ 2020-5),
CAPES (88887.485425/2020-00), and S\~ao Paulo Research Foundation (FAPESP
2018/03211-6, 2020/04624-2). We would like to thank 105 Group Science
(www.105groupscience.com).




\begin{thebibliography}{99}
\bibitem{silvio}
S.L.T. de Souza, A.M. Batista, I.L. Caldas, K.C. Iarosz, J.D. Szezech Jr, 
Dynamics of epidemics: Impact of easing restrictions and control of infection 
spread. Chaos, Solitons \&\ Fractals 142 (2021) 110431. 
\bibitem{automata}
J. Dai, C. Zhai, J. Ai, J. Ma, J. Wang, W. Sun, Modeling the spread of 
epidemics based on cellular automata. Processes 9 (2021) 55. 
\bibitem{vac-ca}
S.H. White, A.M. del Rey, G.R. Sánchez, Modeling epidemics using cellular
automata. Applied Mathematics and Computation 186 (2007) 193-202.
\bibitem{history}
K.A. Glatter, P. Finkelman, History of the Plague: An Ancient Pandemic for the
Age of COVID-19, The American Journal of Medicine 134 (2021), 176-181.
\bibitem{spanish}
T.M. Tumpey, C.F. Basler, P.V. Aguilar, H. Zeng, A. Solórzano, D.E. Swayne, N.J,
Cox, J.M. Katz, J.K. Taubenberger,  P. Palese, A. García-Sastre,
Characterization of the reconstructed 1918 Spanish influenza pandemic virus.
Science 310 (2005) 77-80.
\bibitem{sars}
A.M. Lauren, P. Babak, M.E.J. Newman, D.M. Skowronski, R.C. Brunham, Network
theory and SARS: predicting outbreak diversity. Journal of Theoretical Biology
232 (2005) 71-81.
\bibitem{h1n1}
L. Mao, L. Bian, Spatial–temporal transmission of influenza and its health risks
in an urbanized area. Comput. Environ. Urban Syst. 34 (2012) 204-215.
\bibitem{lancet}
M. Voysey, S.A. Clemens, S. Madhi, L. Weckx, P. Folegatti, P. Aley, B. Angus, 
V. Baillie, S. Barnabas, Q. Bhorat, S. Bibi, C. Briner, P. Cicconi, 
E. Clutterbuck, A. Collins, C. Cutland, T. Darton, K. Dheda, C. Chritina, 
Single-dose administration and the influence of the timing of the booster 
dose on immunogenicity and efficacy of ChAdOx1 nCoV-19 (AZD1222) vaccine: 
a pooled analysis of four randomised trials. The Lancet 397 (2021) 881-891. 
\bibitem{mello}
B.A. Mello, One-way pedestrian traffic is a means of reducing personal 
encounters in epidemics. Frontiers in Physics 8 (2020).
\bibitem{seir-vac1}
C. Balsa, I. Lopes, T. Guarda, J. Rufino, Computational simulation of the
COVID-19 epidemic with the SEIR stochastic model. Computational and Mathematical
Organization Theory (2021). https://doi.org/10.1007/s10588-021-09327-y
\bibitem{sharma21}
N. Sharma, A.K. Verma, A.K. Gupta, Spatial network based model forecasting
transmission and control of COVID-19, Physica A 581(1) (2021), 126223.
\bibitem{seirv}
X. Meng, Z. Cai, S. Si, D. Duan, Analysis of epidemic vaccination strategies 
on heterogeneous networks: Based on SEIRV model and evolutionary game. 
Applied Mathematics and Computation 403 (2021) 126172. 
\bibitem{piccirillo21}
V. Piccirillo, Nonlinear control of infection spread based on a deterministic
SEIR model. Chaos, Solitons \& Fractals 149 (2021) 111051.
\bibitem{amaku21a}
M. Amaku, D.T. Covas, F.A.B. Coutinho, R.S.A. Neto, C. Struchiner, A.
Wilder-Smith, E. Massad, Modelling the test, trace and quarantine strategy to
control the COVID-19 epidemic in the state of S\~ao Paulo, Brazil. Infectious
Disease Modelling 6 (2021) 46-55.
\bibitem{amaku21b}
M. Amaku, D.T. Covas, F.A.B. Coutinho, R.S. Azevedo, E. Massad, Modelling the impact of delaying
vaccination against SARS-CoV-2 assuming unlimited vaccine supply. Theor. Biol.
Med. Model 18 (2021), 14.
\bibitem{rbef}
A.M Batista, S.L.T. de Souza, K.C. Iarosz, A.C.L. Almeida, J.D. Szezech Jr.,
E.C. Gabrick, M. Mugnaine, G.L. dos Santos, I.L. Caldas, Simulation of
deterministic compartmental models for infectious diseases dynamics. Rev. Bras.
Ensino Fis. 43 (2021).

\bibitem{reports}
A. Radulescu, C. Williams, K. Cavanagh, Management strategies in a SEIR 
model of COVID-19 community spread. Scientific Reports 10 (2020) 21256.
\bibitem{seir}
J.M. Carcione,  J.E. Santos, C. Bagaini, J. Ba, A simulation of a 
COVID-19 epidemic based on a deterministic SEIR model. Frontiers in Public
Health 8 (2020) 230.
\bibitem{quanxing}
L. Quan-Xing, J. Zhen, Cellular automata modelling of seirs, Chinese Physics
14 (2005) 1370.
\bibitem{reinfeccao}
E. Malkov, Simulation of coronavirus disease 2019 (COVID-19) sceneries 
with possibility of reinfection. Chaos, Solitons \&\ Fractals 139 (2020) 110296.
\bibitem{heliyon}
P. Wintachai, K. Prathom, Stability of SEIR model related to efficiency of
vaccines for COVID-19 situation. Heliyon 7 (2021) e06812.
\bibitem{seir-vac}
M. Etxeberria-Etxaniz, S. Alonso-Quesada, M. De la Sen, On an SEIR epidemic
model with vaccination of newborns and periodic impulsive vaccination with
eventual on-line adapted vaccination strategies to the varying levels of the
susceptible subpopulation. Applied Sciences 10 (2020) 8296.
\bibitem{wireless}
M. Jadidi, S. Jamshidiha, I. Masroori, P. Moslemi, A. Mohammadi, V. Pourahmadi,
A two-step vaccination technique to limite COVID-19 spread using mobile data.
Sustainable Cities and Society 70 (2021) 102886. 
\bibitem{seir-vac2}
M.A. Safi, A.B. Gumel, Mathematical analysis of a disease transmission model
with quarantine, isolation and an imperfect vaccine. Computers \& Mathematics
with Applications 61 (2011) 3044-3070. 
\bibitem{seir-vac3}
P. Yongzhen, L. Shuping, L. Changguo, S. Chen, The effect of constant and pulse
vaccination on an SIR epidemic model with infectious period. Applied
Mathematical Modelling 35 (2011) 3866-3878. 
\bibitem{nava}
A. Nava, A. Papa, M. Rossi, D. Giuliano, Analytical and cellular automaton 
approach to a generalized SEIR model for infection spread in an open 
crowed space. Physical Review Research 2 (2020) 043379.
\bibitem{two-doses}
M. De la Sen, S. Alonso-Quesada, A. Ibeas, R. Nistal, On a discrete SEIR
epidemic model with two-doses delayed feedback vaccination control on the
Susceptible. Vaccines 9 (2021) 398. 
\bibitem{michele}
M. Mugnaine, E.C. Gabrick, P.R. Protachevicz, K.C. Iarosz, S.L.T. de Souza, A.C.L. Almeida, A.M. Batista, I.L. Caldas, J.D. Szezech Jr, R.L. Viana, Control attenuation and temporary immunity in a cellular automata SEIR epidemic model. Chaos, Solitons \&\ Fractals 155 (2022) 111784.
\bibitem{physica}
N. Sharma, A. K. Gupta, Impact of time delay on the dynamics of SEIR epidemic
model using cellular automata. Physica A 471 (2017) 114-125. 
\bibitem{swarm}
M. Kotyrba, E. Volna, P. Bujok, Unconventional modelling of complex system via
cellular automata and differential evolution. Swarm and Evolutionary Computation
25 (2015) 52-62. 
\bibitem{wolfram2}
S. Wolfram, Cellular automata and complexity: collected papers. 1. ed. Reading,
MA: Addison-Wesley, 1994.
\bibitem{wolfram1}
S. Wolfram, Statistical mechanics of cellular automata. Reviews of Modern
Physics 55 (1983) 601. 
\bibitem{wolfram}
S. Wolfram, Cellular Automata. Los Alamos Science (1983). URL:
\url{https://content.wolfram.com/uploads/sites/34/2020/07/cellul}
\url{ar-automata.pdf}
\bibitem{stc1}
G. Schneckenreither, N. Popper, G. Zauner, F. Breitenecker, Modelling SIR-type
epidemics by ODEs, PDEs, difference equations and cellular automata - A
comparative study. Simulation Modelling Practice and Theory 16 (2008) 1014-1023.
\bibitem{borges2015}
F.S. Borges, E.L. Lameu, A.M. Batista, K.C. Iarosz, M.S. Baptista, R.L. Viana,
Complementary action of chemical and electrical synapses to perception. Physica
A: Statistical Mechanics and its Applications 430 (2015) 236-241.
\bibitem{pramana}
F.S. Borges, P.R. Protachevicz, V. Santos, M.S. Santos, E.C. Gabrick, K.C. Iarosz,
E.L. Lameu, M.S. Baptista, I.L. Caldas, A.M. Batista, Influence of inhibitory
synapses on the criticality of excitable neuronal networks. Indian Academy of
Sciences Conference Series 3 (2020).
\bibitem{bin}
S. Bin, G. Sun, C.C. Chen, Spread of infectious disease modeling and analysis of
different factors on spread of infectious disease based on cellular automata.
International Journal of Environmental Research and Public Health 16 (2019)
4683.
\bibitem{vichniac84}
G.Y. Vichniac, Simulating physics with cellular automata. Physica D 10 (1984)
96-116.  
\bibitem{luca21}
L. Meacci, M. Primicerio, G.C. Buscaglia, Growth of tumours with stem cells:
The effect of crowding and ageing of cells. Physica A 570 (2021) 125841.
\bibitem{viana14}
R.L. Viana, F.S. Borges, K.C. Iarosz, A.M. Batista, S.R. Lopes, I.L.
Caldas, Dynamic range in a neuron network with electrical and chemical synapses.
Communications in Nonlinear Science and Numerical Simulation 19 (2014) 164-172.
\bibitem{santos09}
L.B.L. Santos, M. C. Costa, S.T.R. Pinho, R.F.S. Andrade, F.R. Barreto,
M.G. Teixeira, M.L. Barreto, Periodic forcing in a three-level cellular
automata model for a vector-transmitted disease. Physical Review E 80 (2009)
016102.
\bibitem{blavatska21}
V. Blavatska, Yu. Holovatch, Spreading processes in post-epidemic environments.
Physica A 573 (2021) 125980.  
\bibitem{mikler}
A.R. Mikler, S. Venkatachalam, K. Abbas, Modeling infectious disease using
global stochastic cellular automata. Journal of Biological Systems 13 (2005)
421-439.
\bibitem{cavalcante}
A.L.B. Cavalcante, L.P. de Faria Borges, M.A. da Costa Lemos, M.M. Farias, H.S. Carvalho, Modelling the spread of covid-19 in the capital of Brazil using numerical 
solution and cellular automata. Computational Biology and Chemistry 94 (2021) 107554. 
\bibitem{ilachinski}
A. Ilachinski, Cellular automata: a discrete universe. World Scientific (2001).
DOI: \url{https://doi.org/10.1142/4702}
\bibitem{belik}
V. Belik, T. Geisel, D. Brockmann, Natural human mobility patterns and 
spatial spread of infectious diseases. Physical Review X 1 (2011).
\bibitem{weng}
W.G. Weng, T. Chen, H.Y. Yuan, W.C. Fan, Cellular automaton simulation of
pedestrian counter flow with different walk velocities. Physical Review E 74
(2006) 036102.
\end{thebibliography}
\end{document}